\documentclass[useAMS,usenatbib,usegraphicx]{mn2e}

\def\kms{km s$^{-1}$}
\def\msun{M$_{\sun}$}
\def\aap{A\&A}
\def\apjl{ApJ}
\def\apj{ApJ}
\def\apjs{ApJS}
\def\aj{AJ}
\def\mnras{MNRAS}
\def\nat{Nature}
\def\pasp{PASP}
\def\apss{Ap\&SS}

\title[The Merger Rate of ELM WD Binaries]{The Merger Rate of Extremely Low Mass
White Dwarf Binaries: Links to the Formation of AM CVn Stars and Underluminous
Supernovae\thanks{Based on observations obtained at the MMT Observatory, a joint
facility of the Smithsonian Institution and the University of Arizona.}}

\author[W.\ Brown et al.]{Warren R.\ Brown$^1$,
	Mukremin Kilic$^1$\thanks{\em Spitzer Fellow},
	Carlos Allende Prieto$^{2,3}$,
	and Scott J.\ Kenyon$^1$ \\
	$^1$Smithsonian Astrophysical Observatory, 60 Garden St, Cambridge, MA 02138, USA\\
	$^2$Instituto de Astrof\'{\i}sica de Canarias, E-38205 La Laguna, Tenerife, Spain\\
	$^3$Departamento de Astrof\'{\i}sica, Universidad de La Laguna, E-38206 La Laguna, Tenerife, Spain
	}

\begin{document}

\maketitle
\begin{abstract}

	We study a complete, colour-selected sample of double-degenerate binary
systems containing extremely low mass (ELM) $\le$0.25 \msun\ white dwarfs (WDs).  
We show, for the first time, that Milky Way disk ELM WDs have a merger rate of
approximately $4\times10^{-5}$ yr$^{-1}$ due to gravitational wave radiation.  The
merger end-product depends on the mass ratio of the binary.  The ELM WD systems that
undergo stable mass transfer can account for $\ga$3\% of AM CVn stars.  More
importantly, the ELM WD systems that may detonate merge at a rate comparable to the
estimated rate of underluminous SNe, rare explosions estimated to produce only
$\sim$0.2 \msun\ worth of ejecta.  At least 25\% of our ELM WD sample belong to the
old thick disk and halo components of the Milky Way.  Thus, if merging ELM WD
systems are the progenitors of underluminous SNe, transient surveys must find them
in both elliptical and spiral galaxies.

\end{abstract}


\section{INTRODUCTION}

	We use the first complete, well-defined sample of extremely low mass (ELM)  
$\le$0.25 \msun\ WDs \citep{brown10c} to calculate the space density and merger rate
of these unusual double-degenerate systems.  ELM WDs are noteworthy because 1) they
comprise $<$0.2\% of all spectroscopically confirmed WDs \citep{eisenstein06} and 2)
the Universe is not old enough to produce ELM WDs through single star evolution.  
Radial velocity follow-up of the 12 ELM WDs in the Hypervelocity Star Survey
\citep{brown05, brown07b, brown09a} reveals that they are all binaries with short
(1--14 hr) orbital periods. Remarkably, six of our 12 ELM WDs will merge in less
than a Hubble time. A similar merging WD system \citep[J1257+5428,][]{badenes09}
generated excitement when its companion was thought to be a neutron star
\citep{thompson10}.  Improved spectroscopy shows that the system contains a pair of
WDs \citep{marsh10, kulkarni10}, just like the ELM WD systems studied here.

	Depending on the stability of mass transfer in the Roche lobe overflow 
phase, ELM WD binary mergers produce two possible outcomes.
	ELM WD binaries with mass ratios $\la$0.2 experience stable mass transfer
and form AM Canum Venaticorum (AM CVn) systems, a class of ultracompact binaries
that consist of a WD accretor and a helium donor star \citep{warner95, nelemans05,
solheim10}.  With their short $<$1 hr orbital periods, AM CVn stars are important
gravitational wave sources \citep{hils00, nelemans04}.  Three proposed formation
channels for AM CVn stars involve three different types of donor stars:  WDs, helium
stars, or slightly evolved main-sequence stars \citep[e.g.][]{podsiadlowski03}.  
Distinguishing the dominant formation channel for AM CVn stars is difficult,
however, because all three channels lead to the same helium mass-transfer state.  
Studying the space density and merger rates of progenitors such as our sample of ELM
WD binaries places an important constraint on AM CVn formation.

	ELM WD binaries with mass ratios $\ga$0.7 experience unstable mass transfer
that will result in a merger or possibly an explosion.  Whether ELM WD binaries
merge or explode, and how they do so, are open theoretical problems.  This Letter is
focused on observations:  we compare the observed merger rate of ELM WD binaries
with one possible outcome relevant to current transient surveys, underluminous
supernovae (SNe).

	Underluminous SNe, such as SN~2005E and SN~2008ha, are rare types of
supernova explosions that are 10--100 times less luminous than a normal SN Type Ia
and have only $\sim$0.2 \msun\ worth of ejecta.  One proposed origin for
underluminous SNe is the detonation of a helium layer on a sub-Chandrasekhar mass WD
\citep{nomoto82, woosley86}.  \citet{guillochon10} argue that instabilities in the
accretion stream can detonate the helium layer.  In the ``.Ia'' scenario, a
sufficient mass of helium accumulates in the AM CVn mass transfer phase to ignite
and explode \citep{bildsten07, shen10}.  It is unclear which scenario actually
occurs for a merging ELM WD, however, because the physics of the merger process,
such as the mass transfer rates, are poorly constrained.

	Observationally, ELM WD merger models can successfully explain the stellar
environment, nucleosynthesis products, spectra, and light curves of observed
underluminous SNe.  Half of the known calcium-rich underluminous type Ib/c SNe are
observed in old-population environments \citep{perets10a}:  SN~2005E is located 11
kpc (projected)  above the disk of its host galaxy, while SN~2000ds, SN~2005cz, and
SN~2007ke are located in elliptical galaxies.  These objects cannot be explained by
the core-collapse of massive stars, but are easily explained by merging WDs.  A low
ejecta mass of calcium-rich material is a natural result of burning helium-rich
material on a WD \citep[e.g.][]{woosley86}.  The rapid light curve decay of
SN~2010X, for example, can be explained with the .Ia model \citep{kasliwal10}.
\citet{waldman10} are able to model both the observed spectrum and the light curve
of SN~2005E with the detonation of a 0.2 \msun\ He-layer on a 0.45 \msun\ WD.  Half
of our merging ELM WD systems have similar mass ratios.  Measuring the rate of our
ELM WD mergers thus provides an additional observational constraint on links to
underluminous SNe.

	Our Letter is organised as follows.  In Section 2 we discuss the ELM WD
sample.  In Section 3 we derive the space density of ELM WDs.  In Section 4 we
estimate the merger rate of ELM WDs in the Milky Way, $4\times10^{-5}$ yr$^{-1}$.  
This first estimate suffers from factors of 2 uncertainties, however it demonstrates
that merging ELM WD binaries are viable progenitors of both AM CVn stars and
underluminous SNe.

\section{SAMPLE}

	Our sample of twelve $\le$0.25 \msun\ ELM WDs comes from a radial velocity
survey of stars selected uniformly by magnitude $15<g<20.5$ and colour.  The
approximate colour limits are $0.1<(u-g)_0<0.9$, $-0.45<(g-r)_0<-0.2$, and
$-0.5<(r-i)_0<0$; \citet{brown07b, brown09a} provide the exact prescription.  The
colour-selection was designed to find halo stars with late-B type colours.  However,
15\% of the survey stars are WDs, 12 of which have $\le$0.25 \msun .  Follow-up
spectroscopy reveals that eleven of the 12 ELM WDs are binaries with 1--14 hr
orbital periods \citep{brown10c}.  The single non-variable ELM WD is consistent with
the number of pole-on systems expected in a sample of 12 non-kinematically selected
targets.  Thus we assume that all of the ELM WDs are binaries, and that the orbital
parameters of the 11 ELM WDs fairly sample the underlying distribution.

	We estimate luminosities and distances for our ELM WDs using the updated
\citet{panei07} evolutionary tracks for He-core WDs.  The luminosity estimates
encompass $+8.0<M_g<+9.9$, and correspond to heliocentric distances of 0.3 kpc $<d<$
3.1 kpc \citep{brown10c}.  The sample covers a 9800 deg$^2$ region of the SDSS Data
Release 7 imaging footprint, and is currently 90\% complete in this area.  Thus we
assume that there are 10\% more ELM WDs in our colour-magnitude range when
estimating the ELM WD space density.

\section{DENSITY}

	The simplest approach to estimating the space density of ELM WDs is to
divide the observed number of ELM WDs by the survey volume.  This simple approach
provides a robust lower limit, but only for ELM WDs sampled by our survey.  ELM WDs
are missing from our sample 1) because they cool and 2) because they merge.  Thus we
need to correct our survey for ELM WDs that have cooled outside of our colour range
and that have merged.  Our survey samples ELM WDs that formed in the last 1 Gyr,
thus that is what we consider here.  For a constant formation rate, we will show
that older ELM WDs contribute only a few percent to the present merger rate.  
Because the contribution of these delayed mergers is much smaller than our
uncertainties, the unknown formation history of ELM WDs does not significantly
affect our conclusions.  In this section we estimate the present number of ELM WDs
that exist in the Galaxy.  In the next section we estimate the ELM WD merger rate.

\subsection{Cooling Time Correction}

	Once formed, WDs cool and become redder and fainter with time.  The rate of
cooling depends sensitively on WD mass.  Updated \citet{panei07} evolutionary tracks
for He-core WDs show that 0.17 \msun\ WDs spend $\simeq$1 Gyr cooling through our
colour (effective temperature) range with approximately constant luminosity,
$M_g\simeq+8.0$.  On the other hand, 0.2 -- 0.25 \msun\ WDs spend only $\simeq$0.25
Gyr cooling through our colour range and with declining luminosities, $M_g$ dropping
from +8 to +10.  The difference between the 0.17 \msun\ and 0.2--0.25 \msun\ He-core
WDs is explained by the thermonuclear flash threshold in the hydrogen shell burning
phase \citep{panei07}.  We note that the tracks have the WDs spending $10^7-10^8$ yr
at hotter temperatures, or 1\%--10\% of the time spent in our colour range.  
\citet{serenelli01} tracks yield similar numbers.

	Clearly, our colour selection samples ELM WDs that have formed within the
last 1 Gyr.  In addition, the relation between mass and cooling time skews the
number of ELM WDs of different mass that we observe.  Each observed WD implies the
formation of (1 Gyr)/$t_{obs}$ such objects in the last Gyr, where $t_{obs}$ is the
time spent cooling through our colour range in Gyr.  We take $t_{obs}=1$ Gyr for the
eight $\simeq$0.17 \msun\ WDs in our sample, and $t_{obs}=0.25$ Gyr for the four 0.2
-- 0.25 \msun\ WDs \citep{serenelli01, panei07}.  This correction factor links the
observed sample of ELM WDs to the total number that have formed in the last Gyr that
presently exist in our survey volume.

	The relation between luminosity and temperature also affects the volume over
which we observe different mass ELM WDs.  This is accounted for in our $1/V_{\rm
max}$ density estimate below, with the implicit assumption that the population of
ELM WDs is the same in all volumes.

\subsection{Merger Time Correction}

	ELM WDs are missing from our sample because they merge.  The merger time due
to angular momentum loss from gravitational wave radiation is \[ \tau =
\frac{(M_1 + M_2)^{1/3}}{M_1 M_2} P^{8/3} \times 10^{7} {\rm yr} \label{eqn:tau} \]
where the masses are in \msun\ and the period $P$ is in hours \citep{landau58}.

	Both the mass ratio $M_1/M_2$ and the merger time $\tau$ depend on
the mass of the unseen companion and the unknown orbital inclination ($M_2\sin{i}$).  
An edge-on orbit with $i=90\degr$ gives the minimum possible companion mass and the
longest possible merger time. The average inclination for a random stellar sample is
$i=60\degr$.  We assume that the ELM WDs have the average inclination in the
following discussion and in the values tabulated in Table 1.

	Four of the ELM WDs have merger times less than the time spent cooling
through our colour range.  If the formation rate is constant over the last Gyr, then
the presence of these short merger time objects implies that (1 Gyr)/$\tau$
such objects formed over the last Gyr.  This correction factor links the ELM WDs
that presently exist to the total number that have formed in our survey volume in
the last Gyr.

\begin{table}
 \centering
 \begin{minipage}{80mm}
   \caption{ELM WD Properties}
   \begin{tabular}{@{}lccccc@{}}
   \hline
    ID &  $M_1$   &  $t_{obs}$  & $M_1/M_2^*$ & $\tau^*$ & $N_{cor}$ \\
       & (\msun ) & (Gyr)       &             & (Gyr)            & \\
   \hline
 J0755+4906 & 0.17 & 1    & 0.151 &  0.17 &  5.8  \\
 J0818+3536 & 0.17 & 1    & 0.513 &  7.31 &  1.0 \\
 J0917+4638 & 0.17 & 1    & 0.489 & 30.3  &  1.0 \\
 J0923+3028 & 0.23 & 0.25 & 0.528 &  0.11 &  9.4 \\
 J1053+5200 & 0.20 & 0.25 & 0.614 &  0.13 &  7.6 \\
 J1233+1602 & 0.17 & 1    & 0.142 &  1.69 &  1.0 \\
 J1422+4352 & 0.17 & 1    & 0.309 & 34.7  &  1.0 \\
 J1439+1002 & 0.18 & 0.25 & 0.292 & 44.1  &  1.0 \\
 J1448+1342 & 0.25 & 0.25 & . . . & . . . &  4.0 \\
 J1512+2615 & 0.20 & 0.25 & 0.555 & 140.3 &  4.0 \\
 J1630+2712 & 0.17 & 1    & 0.242 & 12.5  &  1.0 \\
J2119--0018 & 0.17 & 1    & 0.163 &  0.42 &  2.4 \\
   \hline
\end{tabular}
\end{minipage}\\
$^* ~i=60\degr$.
\end{table}

\subsection{Local Space Density}

	We now solve for the total number of ELM WDs that formed throughout Galaxy 
in the last Gyr.  First, we combine the correction factors such that we do not 
over-correct systems that merge faster than they cool.
	\[ N_{cor} = {\rm max}( \frac{1}{t_{obs,i}}, \frac{1}{\tau_i}), \]

	Next, we calculate the local space density of ELM WDs using the modified
$1/V_{\rm max}$ method \citep{schmidt75} described by \citet{roelofs07a}.  This
method accounts for the variation of space density with distance above the Galactic
plane using a disk model.  We use the \citet{roelofs07a} thin- and thick-disk model
combined with the \citet{nelemans01c} exponential radial scale length
	\[ \frac{\rho(Z, R)}{\rho_0} = \left[0.98 ~{\rm sech}( \frac{Z}{0.3}) +0.02 
~{\rm sech}( \frac{Z}{1.25})\right] ~{\rm exp}(\frac{8.5-R}{2.5}), \label{eqn:dens} \]
	where $Z$ is the vertical distance above the Galactic plane in kpc and $R$
is the radial distance along the Galactic plane in kpc.  Although the thick disk
fraction adopted here is low compared to more recent work \citep[e.g.][]{juric08},
this facilitates a fair comparison with the AM CVn population derived by
\citep{roelofs07a}.

	To solve for the local space density of the disk, we exclude halo ELM WDs
from our sample.  Two probable halo objects are J0818+3536 and J1422+4352, which
have systemic radial velocities of $-202$ \kms\ and $-193$ \kms , respectively.  
\citet{kilic10} argue that J1053+5200 is also a halo object based on its proper
motion.  Reliable proper motions are unavailable for the majority of our ELM WDs,
however.  We choose to make a self-consistent cut on systemic radial velocity.  
The inclusion of a spurious halo object in our disk sample would cause us to
over-estimate the local space density by $\simeq$10\%.

	The volume sampled by our survey varies with both radial distance $R$ and
vertical distance $Z$.  We compute $V_{max}$ by starting with the SDSS stripe
definitions, converting them to Galactic coordinates, and then integrating the volume
over our $15<g<20.5$ apparent magnitude range for each WD absolute magnitude.  

	Given the assumed disk model, the completeness-corrected local space density
of disk ELM WDs is $\rho_0=40$ kpc$^{-3}$.  The error is dominated by the factor of
2 uncertainty in our correction factors.  Integrating the disk model over $0<R<25$
kpc and $0<|Z|<10$ kpc, we estimate that $5\times10^4$ ELM WDs formed in the Milky
Way disk in the last Gyr.

\begin{figure}		
 \includegraphics[width=3.25in]{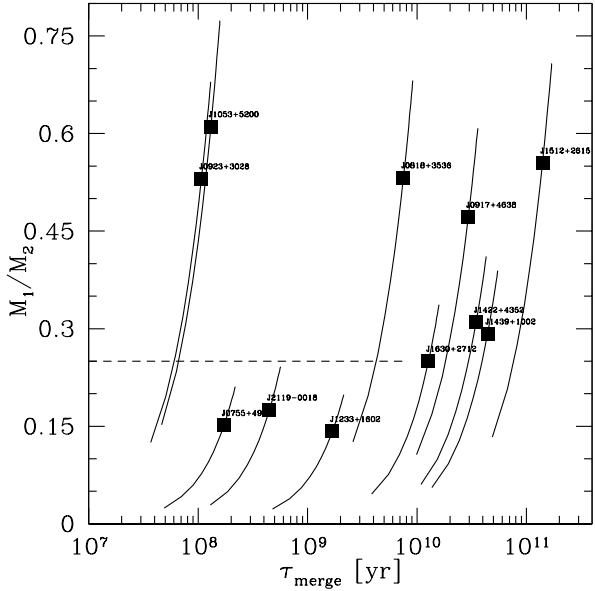}
 \caption{ \label{fig:merge}
	Gravitational wave radiation merger time versus mass ratio for the sample of
ELM WDs, excluding the probable pole-on system. $M_1$ is the observed ELM WD and
$M_2$ is the unseen binary companion.  Filled squares show the most probable values
for $i=60\degr$, and lines indicate the possible range of values over the
90-percentile range $25\degr<i<90\degr$.  ELM WD binaries with mass ratios below the
dashed line will experience stable mass transfer and become AM CVn stars; binaries
above the dashed line will experience unstable mass transfer and may explode as
underluminous SNe. }
 \end{figure}

\section{MERGER RATES}

	The merger times are encoded in the orbital period distribution of the ELM
WDs.  In principle, we can calculate the initial period distribution required to
produce the observed period distribution under the assumption of gravitational wave
radiation energy loss.  Given our small number statistics, this approach is fraught
with uncertainty.

	The approach we take is to consider the observed sample of ELM WDs a
snapshot in time of the whole population.  We use the observed orbital periods as a
proxy for the underlying period distribution.  For reference, Figure \ref{fig:merge}
plots the mass ratio $M_1/M_2$ and merger time $\tau$ for our sample of ELM
WDs, excluding the probable pole-on system.  Filled squares in Figure
\ref{fig:merge} are the most probable values for $i=60\degr$, and lines show the
possible values over the 90-percentile range $25\degr<i<90\degr$.

	Four of the ELM WD systems in our sample have merger times less than 1 Gyr.  
The short merger times of these systems suggest that they represent 70\% of the ELM
WD systems that formed in the last Gyr, as determined by the correction factors
calculated above.  Combining the above numbers, the merger rate of ELM WDs in the
Galaxy is approximately $0.7 \cdot 5\times10^4 / 1$ Gyr $\simeq 4\times10^{-5}$
yr$^{-1}$.  We get a similar number if we divide the number of ELM WDs by the
harmonic mean of the observed merger times.

	A $4\times10^{-5}$ yr$^{-1}$ merger rate is likely an underestimate for a
few reasons.  First, some ELM WD systems formed long ago and are now merging.  In
our corrected sample, 5\% of ELM WDs have mergers times 1--8 Gyr.  If the formation
rate of ELM WD systems was constant over the 8 Gyr age of the disk
\citep{leggett98}, then the contribution of delayed mergers is approximately
$10^{-6}$ yr$^{-1}$.  The contribution increases if we account for the increased
star formation rate at early time, but deceases if we account for the delay to form
ELM WD systems.  Fortunately, the contribution of delayed mergers is so small that
knowing the detailed formation history of ELM WDs is unnecessary; the ELM WD merger
rate is dominated by short merger times.

	Secondly, our observations may be missing ELM WD systems with merger times
$<10^8$ yr.  For a 1 Gyr lifetime in our survey region, we expect ELM WDs with
merger times $<10^8$ yr to contribute $< 10\%$ to our sample.  Objects with $<10^8$
yr merger times have $<1$ hr periods, periods in the realm of observed AM CVn
systems.  The contribution of ELM WD systems formed with $<1$ hr periods may be
better estimated from AM CVn rates.  A single 10$^7$ yr merger would double our
estimated merger rate.

	Finally, accounting for the halo and bulge components of the Milky Way would
also increase the merger rate estimate:  star count models suggest that the stellar
halo contains 10\% of the stellar disk mass \citep{juric08}, and the bulge a similar
amount \citep{widrow05}.  All of these corrections act to increase the merger rate,
thus our estimate is likely a lower limit to the true ELM WD merger rate.

\subsection{Comparison with AM CVn Rates}

	The ELM WD systems with mass ratios $<$0.25 will likely become stable
mass-transfer AM CVn systems.  Based on the mass functions, the companions to these
objects must have $>$0.5 \msun\ and thus are probably normal C/O core WDs.  
Although the stable mass-transfer ELM WDs comprise 50\% of the observed merger
systems (2 of 4 with $\tau<1$ Gyr, or 3 of 6 with $\tau<8$ Gyr), the correction
factors suggest that they account for 33\% of the overall ELM WD merger rate.  The
different percentages illustrate the small number statistics involved.  Thus we
consider the range of rates, $(1-2)\times10^{-5}$ yr$^{-1}$, with the reminder that
our merger rates are likely underestimates.

	\citet{roelofs07a, roelofs07b} make an order-of-magnitude estimate of the
disk AM CVn birth rate of $5\times10^{-4}$ yr$^{-1}$ based on samples containing 5
and 6 AM CVn stars.  Comparing with the stable mass-transfer ELM WDs in our sample
suggests that ELM WDs contribute at least 2\%--4\% of the AM CVn population.  
Although this is an order-of-magnitude comparison, our observations show for the
first time that merging ELM WD systems are a likely source of AM CVn stars.

\subsection{Comparison with Underluminous SNe Rates}

	The ELM WD systems with mass ratios of $\simeq$0.55 will likely become
unstable mass transfer systems.  These systems include the two shortest merger time
objects; the correction factors imply that unstable mass-transfer ELM WDs account
for 67\% of the ELM WD merger rate.
	Given the observed mass ratios, there is a 40\% probability that the ELM WD
companions are C/O WDs (M$\ge$0.45 \msun ).  That means 40\% of the unstable mass
transfer systems may lead to surface He detonation on a C/O WD
\citep[e.g.][]{guillochon10}. Stable-mass transfer systems may also detonate as
``.Ia'' SNe \citep[e.g.][]{bildsten07}.  Thus the rate of ELM WD mergers that may
become underluminous SNe is $(2-3)\times10^{-5}$ yr$^{-1}$, with the reminder that
these rates are likely underestimates.

	For comparison, the Type Ia SNe rate of the Milky Way is $5\times10^{-3}$
yr$^{-1}$ \citep{li10}.  ELM WDs are an unlikely source of Type Ia SNe because of
their low system mass.  The relevant comparison is with underluminous SNe.

	\citet{foley09} estimate that a SN~2008ha-type explosion is observable to a
distance of 40 Mpc, a volume that has had 60 known SNe Type Ia in the past 10 years.  
Thus, based on one object, \citet{foley09} estimate that the underluminous SNe rate
is $\sim$2\% of the Type Ia rate, or $\sim$$1\times10^{-4}$ yr$^{-1}$.  This is a
few times larger than the rate of ELM WD mergers that may become underluminous SNe.  

	We close by noting that at least two of our ELM WDs are halo objects and an
equal or greater number are probably thick disk objects.  The halo and thick disk
are old stellar populations.  If ELM WDs are the progenitors of underluminous SNe,
underluminous SNe must be found in both elliptical and spiral galaxies.

\section{CONCLUSIONS}

	We study the space density and merger rate of double-degenerate binary
systems containing $\le$0.25 \msun\ ELM WDs.  This work is motivated by the first
complete, well-defined sample of 12 ELM WDs fortuitously targeted by the
Hypervelocity Star Survey \citep{brown05, brown07b, brown09a}.  The ELM WDs are all
consistent with being short-period binaries \citep{brown10c}.  Remarkably, 6 of the
11 systems with good period determinations will merge in less than a Hubble time due
to gravitational wave radiation.  

	The merger rate of disk ELM WD binaries in the Milky Way is approximately
$4\times10^{-5}$ yr$^{-1}$.  Although the uncertainty in this estimate is at least a
factor of a few, the observations establish that merging ELM WD binaries are viable
progenitors for several interesting phenomena.  ELM WD systems that undergo stable
mass transfer contribute at least 2\%--4\% of the observed AM CVn population.  More
importantly, ELM WD systems that may detonate merge at a rate comparable to the
observed rate of underluminous SNe.  Other possible outcomes include mergers that
form extreme helium stars or helium-rich sdO stars.

	Further progress requires advances in both theory and observation.  
Understanding how ELM WD binaries merge and ignite is an open problem.  Greater
numbers of AM CVn stars must be found to better constrain their properties
\citep{rau10}.  New imaging surveys such as the Palomar Transient Factory,
Pan-STARRS, Skymapper, and eventually the Large Synoptic Survey Telescope will
discover large numbers of unusual transients such as underluminous SNe.

	Our goal is to use observations to investigate links between
double-degenerate mergers and unusual transients.  In the future, obtaining a larger
sample of 30 ELM WDs will improve the uncertainty in our ELM WD space density and
merger rate estimates by a factor 2.  We anticipate this is the first step in our
ELM Survey towards a greater understanding of merging ELM WD systems.

\section*{Acknowledgements}

	This research makes use of NASA's Astrophysics Data System Bibliographic
Services.  This work was supported in part by the Smithsonian Institution.  MK is
supported by NASA through the {\em Spitzer Space Telescope} Fellowship Program,
under an award from CalTech.


\end{document}